\documentclass[aps,preprint,amsmath,amssymb,groupedaddress,nofootinbib]{revtex4-2}
\usepackage[utf8]{inputenc}
\usepackage{amssymb}
\usepackage{amsmath}
\DeclareMathOperator{\sgn}{sgn}
\usepackage{amsfonts}
\usepackage{mathtools}
\usepackage{bbm}
\usepackage[dvipsnames]{xcolor}
\usepackage{tensor}
\usepackage{booktabs}
\usepackage{physics}
\usepackage{geometry}

\usepackage{tikz}
\usepackage{pgfplots}

\usepackage{tikz-3dplot}

\usetikzlibrary{math}
\pgfplotsset{compat=newest}

\usetikzlibrary{arrows.meta}

\usepackage{amsfonts}
\usepackage{mathtools}
\usepackage{bbm}
\usepackage{xcolor}
\usepackage{tensor}
\usepackage{booktabs}
\usepackage{geometry}
\usepackage[inline]{enumitem}
 
\begin{document}

\title{Non-singular extension of the Kerr-NUT - (anti) de Sitter spacetimes \footnote{We dedicate this paper to the memory of late Ted Newman.}}

\author{Jerzy Lewandowski}
\email{Jerzy.Lewandowski@fuw.edu.pl}
\affiliation{Faculty of Physics, University of Warsaw,\\
ul. Pasteura 5, 02-093 Warsaw, Poland}
\author{Maciej Ossowski}
\email[]{Maciej.Ossowski@fuw.edu.pl}
\affiliation{Faculty of Physics, University of Warsaw,\\
ul. Pasteura 5, 02-093 Warsaw, Poland}

% \thanks{thanks}

\date{\today}

\begin{abstract} 

In 1963 Ezra Ted Newman and his two students Louis A. Tamburino, and Theodore W. J. Unti, proposed a deformation of the Shwarzschild spacetime that made it twisting.
In the cosmological context, an equivalent solution had been found earlier, in 1951, by Abraham Haskel Taub.
The problem that these solutions have is a conical singularity along the  symmetry axis at all distances from the origin.
In 1969 Misner proposed a non-singular interpretation of Taub-NUT spacetimes. 
We extend and refine his method to include a broader family of solutions and completely solve the outstanding issue of a non-singular extension of the Kerr-NUT- (anti) de Sitter solutions to Einstein's equations.
Our approach relies on an observation that in 2 dimensional algebra of Killing vector fields there exist 2 distinguished vector fields that may be used to define $U(1)$-principal bundle structure over the non-singular spaces of non-null orbits.
For all admissible parameters we derive appropriate Killing vector fields and discuss limits to spacetimes with less parameters.
The global structure of spacetime, together with non-singular conformal geometry of the infinities is presented and (possibly also projectively non-singular) Killing horizons is presented.

\end{abstract} 

\maketitle

\tableofcontents

\section{Introduction}
In 1963 Ezra Ted Newman and his two students Louis A. Tamburino, and Theodore W. J. Unti introduced a deformation of the Schwarzschild spacetime that made it twisting.
In the cosmological context, an equivalent solution to Einstein's equation had been found earlier, in 1951, by Abraham Haskel Taub. However, the NUT paper had an enormous impact on the theory of exact solutions to the Einstein equations. 
In particular it led to the Kerr solution \cite{NUT,Kerr}.

The NUT-like modification may be generalized to the Kerr solution to the vacuum Einstein equations by adding a parameter $l$ called the NUT parameter.
The resulting Kerr-NUT spacetime is still Ricci flat, but its topology is considerably different than that of the Kerr spacetime.
Due to  the Misner's method of compactifying the symmetry group \cite{misner}, the global structure of those spacetimes is obtained as $\mathbb{R}\times S^3$ and it contains closed timelike curves.
On the other hand, for sufficiently large value of $l$, the Kerr singularity is smoothed out, although the spacetime still contains horizons.
Due to the unquestionably growing relevance of the cosmological constant in physics, it is natural to generalize the family of the Kerr-NUT solutions by adding a constant $\Lambda$.
That has been done a long time ago. 
These solutions to the vacuum Einstein equations with a cosmological constant are referred to as the Kerr-NUT-(anti) de Sitter spacetimes and set a $4$-dimensional family and since then has been of an interest \cite{Millerdoi:10.1063/1.1666343,Mars2013,Griffiths_2007,griffiths_podolsky_2009,gp2006}.

In the original case of the Taub-NUT spacetime the recipe for Misner's gluing consisted of connecting two patches Taub-NUT spacetime into a non-singular one. This had the consequence of compactifying the orbits of $\partial_t$ Killing vector to circles.
However, in the case of Kerr-NUT-(anti-) de Sitter, a generalized Misner's gluing does not work properly - there still persists a conical singularity irremovable at least from one of the axis of the rotational symmetry.
We completely solve that problem in the current paper, thus our approach generalises the Misner's compactification both in used methods and the class of applicable spacetimes.

We recognize a geometric mechanism of the problem - it is hidden in the spaces of non-null orbits of Killing vectors of Kerr-NUT-(anti) de Sitter spacetimes.
Only some distinguished Killing vectors define a non-singular geometry and we find all of them.
Next, one of those fields is used to perform a non-singular generalization of Misner's gluing. We study the global properties of the resulting spacetime from the past conformal infinity $\mathcal{I}^-$ to the future one $\mathcal{I}^+$, as well as the contained Killing horizons.       

This paper is the third in the series concerning the non-singular interpretation of Kerr-NUT-de Sitter spacetimes.
However, it is a completely self contained continuation.
In the previous papers we focused on the geometry of the Killing horizons contained in that family of spacetimes.
We introduced a notion of a projectively non-singular horizon, i.e. the horizon is said to be projectively non-singular if its space of null generators in non-singular, and derived a $3$-dimensional subfamily of the Kerr-NUT-de Sitter spacetimes, each of which contains a projectively non-singular horizon.
For those Kerr-NUT-de Sitter spacetimes (of a well tuned value of the cosmological constant) we were able to introduce a generalized Misner's construction in a non-singular manner.
This is a special case of the generalization derived in the current paper that is valid for all the values of $\Lambda$ independently of the remaining three parameters, regardless of the projective properties of the horizons.
If a Kerr-NUT-de Sitter spacetime happens to contain a projectively non-singular horizon, then the current construction of non-singular spacetimes may be reduced to the one presented in \cite{LO2}.
Thus the previous results fit neatly into the new ones.      

\section{Kerr-NUT-(anti-) de Sitter spacetimes and our approach}
\subsection{Kerr-NUT-(anti-) de Sitter spacetimes and their problems}
\label{sec:kndsInGeneral}
The Kerr-NUT-(anti-) de Sitter metric in the simplified form first derived by Griffith and Podolsky \cite{Griffiths_2007} can be expressed in the Boyer–Lindquist-like coordinates as

\begin{equation}
\label{eq:KNdS-metric}
\begin{split}
    ds^2=-\frac{\mathcal{Q}}{\Sigma}(dt-A d\phi)^2  +\frac{\Sigma}{\mathcal{Q}}dr^2 
+\frac{\Sigma}{P}d\theta^2+\frac{P}{\Sigma}\sin^2\theta(adt-\rho d\phi)^2,
\end{split}
\end{equation}
where
\begin{equation}
\label{eq:metricFunctions}
    \begin{split}
        \Sigma&=r^2+(l+a\cos\theta)^2,\\
        A&=a\sin^2\theta+4l\sin^2\tfrac{1}{2}\theta,\\%\quad A='a\sin^2\theta-4l\cos^2\tfrac{1}{2}\theta\\
        \rho&=r^2+(l+a)^2=\Sigma+aA,\\% \quad \rho'=r^2+(l-a)^2=\Sigma+aA'\\
        \mathcal{Q}&=(a^2-l^2)-2mr+r^2-\Lambda\big((a^2-l^2)l^2+(\tfrac{1}{3}a^2+2l^2)r^2+\tfrac{1}{3}r^4\big),\\
        P&=1+\frac{4}{3}\Lambda al  \cos\theta+\frac{\Lambda}{3}a^2\cos^2\theta.
    \end{split}
\end{equation}
Above,  $l$ and $a$ denote the NUT and the Kerr parameters, respectively,  $\Lambda$ is a cosmological constant of any value and $m$ stands for the mass parameter (when $\Lambda \neq 0$,  $m$ is proportional to the conserved quantity corresponding to time translation symmetry \cite{Ashtekar_2014}, while for $\Lambda=0$ it is exactly the mass.

Throughout this paper we use a generalisation of the (ingoing) Eddington-Finkelstein coordinates adopted to the rotating spacetime 
\begin{equation}
\label{eq:EddigntonCoords}
    dv:=dt+\frac{\rho}{\mathcal{Q}}dr,\quad
    d\Tilde{\phi}:=d\phi +\frac{a}{\mathcal{Q}}dr.
\end{equation}
This provides an extension of the metric (\ref{eq:KNdS-metric}) that covers the roots of the function ${\mathcal{Q}}$.  
Then the metric tensor takes the following form
\begin{equation}
\label{eq:metricThroughHorizon}
    ds^2=-\frac{\mathcal{Q}}{\Sigma}(dv-A d\Tilde{\phi})^2+2 dr (dv-A d\Tilde{\phi})+\frac{\Sigma}{P}d\theta^2+\frac{P}{\Sigma}\sin^2\theta(a dv-\rho d \Tilde{\phi})^2.
\end{equation}
The above metric shows singularities (apparent or true) familiar from the analysis of the standard Kerr and Kerr-(anti) de Sitter solutions, which are special cases of the considered metrics.

We emphasise now the consequences of the presence of the NUT parameter $l$. A helpful consequence of 
\begin{equation*}
    |l|>|a|
\end{equation*}
is that the function $\Sigma$ never vanishes. Otherwise, if
\begin{equation*}
    |l|\le|a|
\end{equation*}
the function $\Sigma$ takes the value zero at $ r=0$ and  $\theta=\theta_c $ such that
$$ \cos\theta_c = -\frac{l}{a}.$$
That is a source of a  non-removable curvature singularity \cite{griffiths_podolsky_2009}. 
The singularity has a similar structure to that of Kerr, in particular, there are continues curves that pass from the $r>0$ region to the region of $r<0$ such that $\Sigma$ is finite along them, hence they avoid the singular regions.
Therefore, this singularity does not split spacetime into two disconnected components corresponding to $r>0$, and $r<0$, by the analogy to the Kerr singularity.
In the current paper we admit all the values of $a$ and $l$, hence the vanishing $\Sigma$ singularity either appears or not. 

If the NUT parameter $l$ is large enough while $a$ and $\Lambda$ are kept constant the function $P(\theta)$ changes the sign for some $\theta\in [0,\pi]$. That is accompanied by a change of the signature of the metric tensor. To avoid this pathology we allow in the current paper only those values of  $l,a$ and $\Lambda$ that ensure 
\begin{equation}
\label{eq:P>0} P(\theta)>0, \ \ \  {\rm for \ \ every}\ \ \    \theta\in [0,\pi].
\end{equation}
This amounts to
\begin{equation}
\begin{split}
\label{eq:P-ineq}
       P>0 \iff &\Bigg(\left( 2\left|\frac{l}{a}\right|\leq 1 \land 0 < \Lambda < \frac{3}{4l^2}\right) \lor  \left( 2\frac{l}{a} > 1 \land 0< \Lambda < \frac{-3}{a^2-4al}\right) \lor\\ &  \left(-2\frac{l}{a} > 1 \land 0 < \Lambda < \frac{-3}{a^2+4al}\right)
       \lor\left(\frac{-3}{a^2-4al}< \Lambda < 0 \land al<0\right)\lor\\&\left(\frac{-3}{a^2+4al}< \Lambda < 0 \land al>0\right) \Bigg)
\end{split}
\end{equation}
It is conceivable, that this assumption could be carefully relaxed by making the inequality non-sharp,  but this lies beyond the scope of this paper. 

The non-vanishing NUT parameter $l\neq0$, introduces a notable difficulty that eventually has topological consequences.
What is peculiar about this case is the singularity of the differential $1$-form $A d\tilde{\phi}$ that is caused by the term
\begin{equation*}
    4l\sin^2\tfrac{1}{2}\theta d\tilde{\phi}.
\end{equation*}
Indeed, when considered on a sphere parametrised by $(\theta,\tilde{\phi})\in [0,{\pi}]\times [0,2\pi)$, that term is discontinuous at  the pole $\theta=\pi$. That singularity can be cured by introducing another chart that covers the pole $\theta=\pi$. It was  defined  first by Misner  in the Taub-NUT case, i.e. 
$$a=0 = \Lambda$$  
and can be easily  generalised  to   arbitrary values of the parameters $a$ and $\Lambda$.
The Misner charts give rise to the topology $\mathbb{R}\times S^3$ of spacetime, and an action of the U(1) group generated by the Killing vector $\partial_t$ of the metric tensor (\ref{eq:KNdS-metric}) that induces a principal fibre bundle structure
\begin{equation}\label{bundle4}
\mathbb{R}\times S^3 \rightarrow \mathbb{R}\times S^2, 
\end{equation} 
that for each sphere $S^3$, defined by a fixed value of the variable $r$, reduces to
\begin{equation}\label{bundle3}
S^3 \rightarrow  S^2.
\end{equation}

In the case of $\Lambda=0$ the glueing solves the problem at $\theta=\pi$.
However, at a generic case of 
$$a l \Lambda \not=0,$$ 
one more obstacle  appears. As long as the fibres of the bundle  (\ref{bundle3}) contained in the spacetime $\mathbb{R}\times S^3$ are not null and the spacetime is twice differentiable, the geometry induced on $S^2$ should be continues and differentiable. The latter one is defined by the angular part of the spacetime metric tensor (\ref{eq:metricThroughHorizon}) (or equivalently of the metric tensor (\ref{eq:KNdS-metric}) except for the horizons)
\begin{equation}
\label{eq:onS2}
    \frac{\Sigma}{P}d\theta^2+\frac{P}{\Sigma}\sin^2\theta \rho^2 d\tilde{\phi}^2
\end{equation}
while the remaining parts are differentiable on $S^2$ on their own and the term $ (4l\sin^2\tfrac{1}{2}\theta d\tilde{\phi})^2 $
is cured by the Misner glueing. It is easy to see \cite{LO2}, that the tensor (\ref{eq:onS2}) gives rise to a well defined and differentiable  metric tensor on entire $S^2$ including  the poles $\theta=0$ and $\theta=\pi$ by a suitable rescaling of the variable $\tilde{\phi}$, if and only if 
\begin{equation}
\label{eq:p0=pPi}
    P(0) = P(\pi),     
\end{equation}
that is the case (see definitions \ref{eq:KNdS-metric} and \ref{eq:metricFunctions}), if and only if  
$$a l \Lambda=0.$$
Otherwise, the angular part of the metric tensor (\ref{eq:onS2}) has an irremovable conical singularity at least at one of the poles.
In the spacetime $\mathbb{R}\times S^3$, the corresponding singularity takes the form of a $2$-dimensional surface on which the variable $r$ takes all the values in $\mathbb{R}$.
Hence, in the case $a l \Lambda \not=0$, the metric tensor  (\ref{eq:KNdS-metric}) can not be extended to an analytic metric tensor  defined on $\mathbb{R}\times S^3$ such that the Killing vector $\partial_t$ generates the fibres of the projection (\ref{bundle4}).
The above naive approach to the solution of the problem is further justified in the following chapters using the broader geometrical picture of the spaces with NUT parameters.

In the current paper we solve the problem of a non-singular generalisation of  Misner's gluing to a general case of the Kerr-NUT-(anti)-de-Sitter spacetime.
The resulting spacetime still has the U(2)-bundle structure (\ref{bundle4}) and the only possible singularities corresponding to zeros of the function $\Sigma$ if $|l| < |a|$.
Otherwise the spacetime is completely singularity free.
% {\color{blue}
Thus adding large enough NUT parameter may be seen as a complete and smooth (as will be demonstrated later) regularization of Schwarzschild and Kerr solutions, which also happens to satisfy Einstein Equations.
% }

The metric tensor is well defined and analytic in the following range of the variables 
\begin{equation*}
   -\infty <v,r< \infty,\ \ 0\leq \theta < \pi,\ 0\leq \tilde{\phi} < 2\pi P(0), 
\end{equation*}
while we still have to take care of the half-axis $\theta=\pi$. 

\subsection{Our approach to the problem}  
{\color{black} Throughout this paper we use the convention that objects with bar (e.g. $\bar{g}$, $\bar{q}$, $\bar{\xi}$, $\bar{\omega}$) are globally defined on the whole $\mathbb{R}\times S^3 $.}
We wish to construct a spacetime metric $\bar{g}$, such  that  $\bar{g}$:
\begin{itemize}
 \item   is locally isometric to (\ref{eq:metricThroughHorizon}) with $l \neq 0$, 
 \item admits an isometric action of the group U(1) that induces the principal fibre structure                     
\begin{equation}
\label{eq:bundle4'}
\Pi : \mathbb{R}\times S^3 \rightarrow \mathbb{R}\times S^2. 
\end{equation} 
\end{itemize}
We start  with addressing necessary conditions for a  choice of a Killing vector $\xi$ of (\ref{eq:metricThroughHorizon}) whose orbits will be compactified. 
If the desired metric $\bar{g}$ exists then, the extension $\bar\xi$ of the vector field $\xi$ is the  generator of the  U(1) action, and as such has to be a nowhere vanishing Killing vector field.
The projection (\ref{eq:bundle4'}) induces a metric tensor $\bar{q}$ on an open subset  
$$ U \subset \mathbb{R}\times S^2,$$ 
of the  non-null (with respect to $\bar{g}$) orbits of the action of U(1) in $\mathbb{R}\times S^3$.
In  every point of $U$ the metric tensor $\bar{q}$ should be well defined and (at least) differentiable.
Therefore, in Sec. \ref{sec:non-singularOrbits} we study the  spaces of orbits of Killing vectors of the form
$$ \partial_v+b\partial_{\tilde{\phi}}$$
in the spacetime  (\ref{eq:metricThroughHorizon}). 
We determine those Killing vectors that define a non-singular metric tensor $q$ on $U$. 
It turns out, that in the presence of $l\not=0$ there are allowed exactly two values of the parameter $f$ for each triple of values of $a,\Lambda$ and $l$. 
As far as we know this consequence of the presence of the NUT parameter $l\neq0$ has not been described in the literature before.
Indeed, in the  $l=0$ case for every value of the parameter $b$, the corresponding Killing vector field  defines a non-singular geometry $q$ on the space of non-null orbits and the problem becomes trivial.

The element of the desired metric tensor $\bar{g}$ on $\mathbb{R}\times S^3$ encoding the  non-trivial structure of the bundle (\ref{eq:bundle4'}) is the  $1$-form of the rotation-connection of the Killing vector $\bar{\xi}$, namely
 \begin{equation}
 \label{eq:omega}
  \bar{\omega} := \frac{\bar{\xi}_\mu dx^\mu}{\bar{\xi}_\alpha \bar{\xi}^\alpha} ,
   \end{equation}
valid wherever 
$$\bar{\xi}_\alpha \bar{\xi}^\alpha\not=0.$$
If the bundle extension (\ref{eq:bundle4'}) of the spacetime (\ref{eq:metricThroughHorizon}) exists, then the part of the spacetime  $\mathbb{R}\times S^3$ described by (\ref{eq:metricThroughHorizon}) is a trivialisation of  (\ref{eq:bundle4'}) that covers the pole $\theta=0$ of $S^2$. 
Therefore,  in Sec. \ref{sec:misnerglueing}, for each of the  Killing vectors $\xi$ derived in Sec. \ref{sec:non-singularOrbits} we derive the rotation-connection $1$-form  $\omega$.
The analysis of the discontinuity  of  $\omega$  as $\theta\rightarrow \pi$ leads to a complementary trivialisation of (\ref{eq:bundle4'}) that covers the pole $\theta=\pi$.
Remarkably, the key limit properties of $\omega$ at $\theta=\pi$ are independent of $r$.
Hence, the second trivialisation covers also the null orbits.
The transformation law between the trivialisations becomes a recipe for bundle reconstruction implemented in \mbox{Sec. \ref{sec:globalstructure}}.
The trivialisations come with metric tensors $g$ and $g'$, respectively.
The former one is the original metric tensor (\ref{eq:metricThroughHorizon}), and the latter is a new, transformed metric.
On the overlap of the trivialisations the metric tensors $g$ and $g'$ are consistent with each other according to the trivialisation transformations.
In that way they consistently make up a uniquely defined metric tensor $\bar{g}$ on the entire manifold $\mathbb{R}\times S^3$ that satisfies all of the desired properties.

\section{Non-singular space of Killing orbits and non-singular KN(a)dS spacetime}
\label{sec:non-singularOrbits}
In this section we consider the geometries of the spaces of orbits of the Killing vector fields \cite{chrusciel} in the spacetime (\ref{eq:metricThroughHorizon}).
In a case of generic parameters , the most general form of nowhere vanishing Killing vector field is   
\begin{equation}
\label{eq:genericKilling}
    \xi=\partial_{v}+b\partial_{\tilde{\phi}},\ \ \ \ \ \ \ b=\rm const.
\end{equation}
For $a=0$, i.r. in the Taub-NUT-(anti-)de Sitter case the metric has richer symmetry group generated by $\mathfrak{su}(2)\oplus\mathfrak{u}(1)$, where the summands correspond right action by a 3D rotations and right action by a time translation, respectively \footnote{In fact every Ricci flat spacetime admitting $SU(2)\times U(1)$ isometry group corresponds to a generalised Taub-NUT spacetime with the topology of $\mathbb{R}$ $\times$ $L(n,1) $, where $L(n,1)$ is a lens space \cite{Taub-NUTLens}}.
Consequently instead of the combination (\ref{eq:genericKilling}) we could equivalently consider its rotation.

In an adapted  coordinates system, that is
\begin{equation}
\label{eq:killingCoords}
  (x^\mu)=(\tau,x^i) =   (v, \ r,\ \theta,\ \hat{\phi}:=-b v+ \Tilde{\phi})
\end{equation}
the Killing vector field $\xi$ takes a simple form
$$ \xi = \partial_\tau.  $$
The three coordinates
$$(x^i)=(r,\theta,\hat{\phi})$$ 
are adopted in the sense that they satisfy
\begin{equation*}
    \xi(x^i) = 0,
\end{equation*} 
hence they set a coordinate system on the space of the orbits.
To find the metric tensor induced thereon, we  use the rotation-connection  $1$-form 
 \begin{equation}
  \omega := \frac{\xi_\mu dx^\mu}{\xi_\alpha \xi^\alpha} ,
   \end{equation}
and decompose the spacetime metric (\ref{eq:metricThroughHorizon}) in the following manner
 \begin{equation}
 \label{eq:decomp}
 g =  \xi^\alpha\xi_\alpha\omega^2  + q.
 \end{equation}
Then, the part $q$  of the spacetime metric  satisfies
 $$ \xi^\mu q_{\mu\nu}=0={\cal L}_\xi q_{\mu\nu}, $$
 hence, it is expressed purely in terms of the three coordinates $(x^i)$,
\begin{equation}   
\label{eq:q1}
q= q_{ij}(r,\theta,\hat{\phi})dx^i dx^j. 
\end{equation}
As a matter of fact, $q$ is the pullback to the spacetime of the metric tensor induced on the space of the non-null orbits of $\xi$. When the variables $r,\theta$, and $\hat{\phi}$ denote both the spacetime and the space of the orbits coordinates, the pullback of the metric on the space of the orbits is given exactly by (\ref{eq:q1}).
We calculated $q$  for the metric tensor  (\ref{eq:metricThroughHorizon})  transformed to the adapted coordinate system  (\ref{eq:killingCoords}), and the result reads as
 \begin{equation}
     \begin{split}
         q=&\frac{(1-A b)^2 \Sigma}{
   \mathcal{Q} (1-A b)^2-P\sin ^2\theta(a-b \rho
   )^2} dr^2 + \frac{P \sin ^2\theta \Sigma (b \rho-a
   )}{\mathcal{Q}
   (1-A b)^2-P \sin ^2\theta (a-b \rho )^2} 2 dr d\hat{\phi}+ \\
   +& \frac{\Sigma }{P} d\theta^2 + \frac{P \mathcal{Q} \sin ^2\theta \Sigma}{\mathcal{Q} (1-A b)^2-P \sin ^2\theta (a-b \rho
   )^2} d\hat{\phi}^2.
     \end{split}
 \end{equation}

The above metric is well defined for $(r,\theta,\hat{\phi}) \in \mathbb{R}\times ]0,\pi[ \times [0,2\pi c[$, as long as the denominators do not vanish.
The parameter $c$ represents the rescaling freedom that will be used to fix the metric at the poles.
The function $P$ is positive everywhere by the assumption, and the other denominators are proportional to  $\xi^\mu\xi_\mu$. More precisely 
\begin{equation}\label{xixi}
   g(\xi,\xi)=g_{\tau\tau}= \Sigma^{-1}\left(P \sin ^2\theta  (a-b \rho )^2-\mathcal{Q} (1-A b)^2\right).
\end{equation}
For general values of the  parameters $m,a,\Lambda,l,b$  it does vanish for some values of $(r,\theta)$,where we do not expect the metric $q$ to be well defined. Hence we consider the metric $q$ only where
\begin{equation}
 P \sin ^2 \theta  (a-b \rho )^2-\mathcal{Q} (1-A b)^2\ \not=\ 0.
\end{equation}
 
The degeneracies of $q$ that we do worry about are the half-axis $p_0$ and $p_\pi$ corresponding to $\theta=0$ and $\theta=\pi$, respectively.
The term proportional to $dr^2$ is manifestly regular, so is the term proportional to $dr\,d\hat{\phi}$ because it can be viewed as a regular  $1$-form  $\sin^2\theta d\hat{\phi}$  times an analytic function times $dr$.
Now we turn to the purely angular part and consider the pullbacks of $q$ to the surfaces of $r$=const, that is
\begin{equation}
\label{eq:q2}
    {}^{(2)}q=\frac{\Sigma }{P} d\theta^2 + \frac{P \mathcal{Q} \sin ^2\theta \Sigma}{\mathcal{Q} (1-A b)^2-P \sin ^2\theta (a-b \rho
   )^2} d\hat{\phi}^2.
\end{equation}
One of the tools we use for the analysis are closed curves of $\theta=\theta_0$, which can be view as circles around either pole (notice, that the $\hat{\phi}=$const curves are geodesic with respect to $ {}^{(2)}q$).
The radii as seen from either pole ($R_0$ or $R_\pi$, respectively) and circumference ($L_0$) are defined as
\begin{equation}
\begin{split}
    R_0(\theta_0)=\int_0^{\theta_0}\sqrt{{}^{(2)}q{}_{\theta\theta}} d\theta, \quad R_\pi(\theta_0)=\int^\pi_{\theta_0}\sqrt{{}^{(2)}q{}_{\theta\theta}} d\theta, \quad
    &L(\theta_0)=\int_0^{2\pi c}\sqrt{{}^{(2)}q{}_{\hat{\phi}\hat{\phi}}} d\hat{\phi}.
    \end{split}
\end{equation} 
Then the condition for removing the conical singularity is recovering the expected limit of $2\pi$ of ratio of the circumference to radius of the aforementioned cures as we tend to the poles
\begin{equation}
\label{eq:theconstraint}
\lim_{\theta_0\rightarrow 0}\frac{L(\theta_0)}{R_0(\theta_0)} = 2\pi = \lim_{\theta_0\rightarrow \pi}\frac{L(\theta_0)}{R_\pi(\theta_0)}.
\end{equation} 
The above amounts to \footnote{One may also consider an extension of the Kerr-NUT-(anti-) de Sitter spacetimes to the case with the acceleration parameter. Then the condition (\ref{eq:theconstraint}) is formally the same, although with more complicated function $P$. See \cite{LO} for a discussion of the non-singularity if $\xi$ develops the horizon.}
\begin{equation}
\label{eq:p0=pPigeneric}
    P(0)=\frac{P(\pi)}{| 1 - 4 l b |}, \quad c=1/P(0).
\end{equation}
We would be in trouble, if this condition involved the coordinate $r$,  however,  this is not the case because the function $P$ depends only on $\theta$.  

Due of the absolute value in the denominator above, for $l\not=0$, there are 2 possible branches of solutions, each depending on the parameters of the spacetime. For the further convenience let us denote 
\begin{equation*}
    \sigma:=\sgn{(1-4lb)}.
\end{equation*}
Either we have $\sigma=1$  and then we find the solution $b = b_+$, such that
\begin{equation}
\label{eq:b+}
0 < 1- 4l b_+ = \frac{P(\pi)}{P(0)}, \ \ \ \ \ \ b_+=\frac{2a\Lambda}{3+a^2\Lambda+4al\Lambda}%\iff \Lambda=\frac{3f}{2a-af(a+4l)},
\end{equation}
or $\sigma=-1$  in which case the solution $f=f_-$, satisfies
\begin{equation}
\label{eq:b-}
0 < 4l b_- -1  = \frac{P(\pi)}{P(0)},  \ \ \ \ \ \ b_-=\frac{3+a^2\Lambda}{2l(3+a^2\Lambda+4al \Lambda)} %\iff    \Lambda=\frac{3(1-2fl)}{a(8fl^2-a(1-2fl))}.
\end{equation}
We note, that the assumed inequalities rewritten  in (\ref{eq:b+}) and  (\ref{eq:b-}) are consistent with the overall assumption that the function $P$ does not vanish for $\theta\in[0,\pi]$.

In either case, the rescaled angle variable  ranging from $0$ to $2\pi$ is
\begin{equation}
\label{eq:varphi}
    \varphi= P(0)\,\hat{\phi} .
\end{equation}

It is instructing to test our results on the special cases that are encountered in the literature. 
A very special  case  when the Killing vector field $\xi$  develops a horizon  was studied extensively in \cite{LO2} and \cite{LO}. Then the coefficient $b$ is related to the value $r_0$ taken by the coordinate $r$ at the horizon, namely 
\begin{equation}
\label{eq:bhor}
 b=\frac{a}{r_0^2+(a+l)^2}.
\end{equation}
Theh is follows that
\begin{equation}
 1-4lb =1 - \tfrac{4la}{r_0^2+(a+l)^2} = \frac{r^2_0+ (a-l)^2}{r_0^2+(a+l)^2}>0,
\end{equation}
hence this choice falls in the very case (\ref{eq:b+}).
The conditions (\ref{eq:bhor}) and (\ref{eq:b+}) determine the value of $\Lambda$, namely
\begin{equation}
\label{eq:projectivehor}
 \Lambda = \frac{3}{a^2+2l^2+2r^2_0}.
\end{equation}
That is exactly the value found in \cite{LO} when the horizon can be made  projectively non-singular, i.e. its space of the null generators is non-singular.
This can be done using the same rescaled coordinate as for the surrounding spacetime.
The horizon is then necessarily cosmological (more precisely: the outermost, possibly with a negative mass parameter) and non-extremal.

Another compelling choice of the Killing vector is simply
\begin{equation*}
    \xi=\partial_v, \ \ \ \ \ \ {\rm meaning}\ \ \ \ \ \ b=0,
\end{equation*}
resembling the original Misner's choice is his non-singular interpretation of the Taub-NUT metric tensor. Upon this choice, the condition (\ref{eq:p0=pPigeneric})   amounts to the constraint 
$$P(0)=P(\pi),$$ 
which is met iff $\Lambda a l = 0$.
We have discussed that case in Sec \ref{sec:kndsInGeneral}.
However now, in view of our general result derived in this section, the value $b=0$ falls into the $b_+$ case (\ref{eq:b+}), while there is yet another solution, the one of the $b_-$ type (\ref{eq:b-}) , namely
\begin{equation*}
    b = \frac{1}{2l}.
\end{equation*}
 The corresponding  the Killing vector field is
 \begin{equation}
     \xi = \partial_v + \frac{1}{2l}\partial_{\tilde{\phi}}.
 \end{equation}
 
Finally, when the NUT parameter is switched off, that is when $l=0$, then every Killing vector field
 $$ \xi = \partial_v + b \partial_{\tilde{\phi}}, \ \ \ \ \ \ \ b=\rm const $$
defines a non-singular geometry on the orbit space wherever $\xi^\mu\xi_\mu \not= 0$.
That is why we never encounter that issue while considering spacetimes without the NUT parameter.    
\bigskip

% \color{violet}
\noindent{\bf Remark.}\  {\it An intriguing and useful observation are the following general identities:}
$$ b_+ + b_- = \frac{1}{2l},$$ 
and
\begin{equation}\label{eq:b-b}
b_- - b_+ = \frac{P(\pi)}{2l P(0)}.  
\end{equation}

% {\color{blue} 
Although from the conceptual point of view satisfying the constraint (\ref{eq:theconstraint}) guarantees only the continuity of the metric, we also recover that the metric is smooth.
The suspicious parts of the decomposition (\ref{eq:decomp}) are connection 1-form $\omega$ and the orbit metric ${}^{(2)}q$ induced on sphere $r=\text{const}$.
The explicit formulae for the before-mentioned tensors are given by (\ref{eq:q2}) after the substitution (\ref{eq:varphi}) and (\ref{eq:omegaExplicit}).
Using the coordinates corresponding to orthogonal projections of a hemisphere covering one of the poles to the plane, we check by inspection that the components of those tensors are smooth.
The details of this procedure are analogous to those described in the Appendix of \cite{LO}.
% }

\section{Generalisation of Misner's  gluing}
\label{sec:misnerglueing}
The starting point for this section is one of the Killing vector fields $\xi$ (\ref{eq:genericKilling}) found in the previous section, that is such that the constant $b$ satisfies one of the conditions: either (\ref{eq:b+}) or  (\ref{eq:b-}).
In terms of the adapted coordinates  (\ref{eq:killingCoords}) with the rescaled angle variable (\ref{eq:varphi}) the metric tensor (\ref{eq:metricThroughHorizon}) takes the following form 
\begin{equation}
\label{eq:metricThroughHorizonKillingCoords0}
\begin{split}
    ds^2=&-\frac{\mathcal{Q}}{\Sigma}((1-b A)d\tau-\frac{A}{P(0)} d\varphi)^2+2 dr ((1-b A)d\tau-\frac{A}{P(0)} d\varphi)+\frac{\Sigma}{P}d\theta^2+\\
    &+\frac{P}{\Sigma}\sin^2\theta((a-\rho b) d\tau-\frac{\rho}{P(0)} d\varphi)^2,
\end{split}
\end{equation}
and the Killing vector field is simply
\begin{equation}
\xi = \partial_\tau .
\end{equation}
If this spacetime is a trivialisation of the principal fibre bundle (\ref{eq:bundle4'}) and $\xi$ is a generator of the structure group action, as we want it to be, then the orbits of $\xi$ are closed curves, and the parameter  $\tau$ takes values in a finite interval 
\begin{equation*}
    \tau\in[0,\tau_0).
\end{equation*}
The relation of $\tau_0$ and the NUT parameter will follow as a consistency condition for a transformation between the given one and a new, complementary trivialisation that will cover the half-axis $\theta=\pi$.
It is the rotation-connection $1$-form (\ref{eq:omega}) that will tell us, how to construct this complementary trivialisation. 
The explicit formula for $\omega$ reads           
\begin{equation}
\label{eq:omegaExplicit}
         \omega=d\tau -\frac{(1-A b) \Sigma dr +\left(A \mathcal{Q} (1-A b)-P   \sin ^2\theta \rho(a-b \rho
   )\right)  d \varphi/P(0)}{\mathcal{Q}
   (1-A b)^2-P \sin ^2\theta (a-b
   \rho )^2}.%\frac{1}{P(0)}.
\end{equation}
It is well defined at $\theta=0$, however it fails to be so at $\theta=\pi$,
\begin{equation}
    \omega_\varphi(r,\theta=0)=0, \quad \omega_\varphi(r,\theta=\pi)=-\frac{4l}{1-4l b}\frac{1}{P(0)}=\sigma\frac{-4l}{P(\pi)}.
\end{equation}
The obstruction  is non-vanishing of the component $\omega_\varphi$ at the second half-axis. 

Along with $\omega$ the metric tensor is not well defined at $\theta=\pi$, what can be seen  from the formula (\ref{eq:decomp}).
From the limit of $\omega_\varphi$ at $\theta=\pi$, we  deduce a coordinate transformation that cures $\omega$ at that half-axis at the cost of $\theta=0$, namely 
\begin{equation}
\label{eq:t'}
\tau=\tau' + \sigma\frac{4l}{P(\pi)}\varphi' , \ \ \ r=r', \ \ \ \theta=\theta', \ \ \ \varphi=\varphi'\ \ \ \ \ {\rm for}, \ \ \ \theta,\theta'\not= 0,\pi. 
\end{equation}
The condition for the constant $\tau_0$ is hidden behind the transformation of $\tau$. If $\varphi$ and $\varphi + 2\pi$  correspond to a same point of spacetime  for every value of $\tau$, $r$, $\theta$ and $\phi$, and the same is true for $\varphi'$ and $\varphi'+2\pi$, also $\tau$ and $\tau'$  must parametrise circles with the period
\begin{equation}\label{tau_0}
\tau_0=2\pi \frac{4l}{1-4l b}\frac{1}{P(0)}=2\pi \sigma\frac{4l}{P(\pi)},
\end{equation}
alternatively the period may be a $\tfrac{1}{n}$ fraction of the above.  
Hence, the coordinates $(\tau,r,\theta,\varphi)$ parametrise $S_1\times \mathbb{R}\times \left(S^2\setminus \{p_\pi\}\right)$, and the coordinates $(\tau',r',\theta',\varphi')$ parametrise  $S_1\times \mathbb{R}\times \left(S^2\setminus \{p_0\}\right)$, where $p_0$ and $p_\pi$ are the poles of $S^2$ corresponding to $\theta=0$ and $\theta=\pi$, respectively. The transformation  (\ref{eq:t'}) defines gluing of the patches and the vector fields $\partial_\tau$ and $\partial_{\tau'}$ give rise to a uniquely defined vector field
\begin{equation*}
    \partial_{\tau}=\bar{\xi}=\partial_{\tau'}.
\end{equation*}
The manifold defined by the two charts is diffeomorphic to $\mathbb{R}\times S^3$ and the flow of $\bar{\xi}$  makes it  the bundle (\ref{eq:bundle4'}).  
The transformation (\ref{eq:t'}) maps the $1$-form $\omega$ into $\omega'$, which is extendable by the continuity to $\theta'=\pi$.
It is analytic $1$-form in the subset of the second chart corresponding to $\xi'^\mu\xi'_\nu\not=0$.
Finally, the $2$-metric tensor $q$ is invariant with respect to the transformation (\ref{eq:t'}).       

Applying the transformation (\ref{eq:t'}) to the metric tensor (\ref{eq:metricThroughHorizonKillingCoords0}) we obtain a metric, which is well defined on the chart containing the pole $\theta=\pi$
\begin{equation}
\label{eq:metricThroughHorizonKillingCoordsPi}
\begin{split}
    ds'^2=&-\frac{\mathcal{Q}}{\Sigma}((1-b A)d\tau'-\frac{A'}{P(\pi)} \sigma d\varphi')^2+2 dr' ((1-b A)d\tau'-\frac{A'}{P(\pi)} d\varphi')+\frac{\Sigma}{P}d\theta'^2+\\
    &+\frac{P}{\Sigma}\sin^2\theta'((a-\rho b) d\tau'-\frac{\rho'}{P(\pi)} \sigma d\varphi')^2,
    \end{split}
\end{equation}
where $A'(\theta') := a\sin^2\theta'-4l\cos^2\tfrac{1}{2}\theta',\ \rho'(r'):=r'^2 +(a-l)^2$.
Note that $A'd\varphi'$ is dual to $Ad\varphi$ is the sense that $A'd\varphi'$ vanishes at $\theta=\pi$ and is singular at $\theta=0$. Another way of discovering these functions would be, instead of starting with metric (\ref{eq:KNdS-metric}) well defined at $p_0$, to start with a metric well defined at $p_\pi$. This can be achieved by a transformation $t'=t+ 4l \phi$ and replacing $A$ and $\rho$ with $A'$ and $\rho'$ in (\ref{eq:KNdS-metric}) and (\ref{eq:metricFunctions}).

Finally, we can turn to the non-singularity of the resulting metric tensor $\bar{g}$.
This issue amounts to showing the non-singularity of the metric tensors (\ref{eq:metricThroughHorizonKillingCoords0}) and (\ref{eq:metricThroughHorizonKillingCoordsPi}) in their charts.
By construction, each of the metric tensors is automatically non-singular as long as 
\begin{equation}
\label{eq:xinormnonzero} 
\bar{\xi}^\mu\bar{\xi}_\mu\not=0,
\end{equation}
owing to the decomposition 
 \begin{equation}
 \label{eq:decomp2}
 g =  \xi^\alpha\xi_\alpha\omega^2  + q, \quad  g' =  \xi'^\alpha\xi'_\alpha\omega'^2  + q
 \end{equation}
 and the non-singularity of $\xi^\alpha\xi_\alpha$, $\xi'^\alpha\xi'_\alpha$, $\omega$, $\omega'$ and $q$ in the corresponding charts.
Notice, that the missing prime at the second $q$ is intentional - indeed, at that stage of the construction we use the single $3$-metric tensor $q$, the same for each chart.

As was argued in the previous section the above components are smooth and thus metric $g$ is smooth whenever the decomposition (\ref{eq:decomp2}) is valid.
 To also cover the surfaces of $\bar{\xi}^\mu\bar{\xi}_\mu=0$ we repeat the procedure used for the analysis of the smoothness of the decomposition.
 Given the metric tensors in the form (\ref{eq:metricThroughHorizonKillingCoords0}) and  (\ref{eq:metricThroughHorizonKillingCoordsPi}) we relax the assumption (\ref{eq:xinormnonzero}) and decompose the formulae into another set of non-singular elements.
%  
% {\color{red} But given the metric tensors in the form (\ref{eq:metricThroughHorizonKillingCoords0}) and  (\ref{eq:metricThroughHorizonKillingCoordsPi}) we relax the assumption (\ref{eq:xinormnonzero}) and decompose the formulae in non-singular elements.}
First of all, except for the half-axes $\theta=0$ and $\theta'=\pi$, all of the coefficients are non-singular.
To analyse the metrics at the poles we decompose them into the following way.
First consider the purely angular parts 
\begin{align} 
&\frac{\Sigma}{P}d\theta^2+\frac{P}{\Sigma}\sin^2\theta\left(\frac{\rho}{P(0)}\right)^2 d\varphi^2 =\frac{\Sigma}{P}\left(d\theta^2+\frac{P^2}{\Sigma^2}\sin^2\theta\left(\frac{\rho}{P(0)}\right)^2 d\varphi^2\right), \ \ \ {\rm at}\ \ \theta=0,\\
&\frac{\Sigma}{P}d\theta'^2+\frac{P}{\Sigma}\sin^2\theta'\left(\frac{\rho'}{P(\pi)}\right)^2 d\varphi'^2=\frac{\Sigma}{P}\left(d\theta'^2+\frac{P^2}{\Sigma^2}\sin^2\theta'\left(\frac{\rho'}{P(\pi)} \right)^2d\varphi'^2\right)\ \ \ {\rm at}\ \ \theta'=\pi.
\end{align}
Employing again the orthogonal projection (as described in Appendix A of \cite{LO} one can see that in the parentheses, the coefficients at $\sin^2\theta d\varphi^2$ and  $\sin^2\theta' d\varphi'^2$ are smooth (analytic) and tend to $1$ at $\theta=0$ and, $\theta'=\pi$, respectively.     

Next, consider the differential $1$-forms appearing in  (\ref{eq:metricThroughHorizonKillingCoords0}) and  (\ref{eq:metricThroughHorizonKillingCoordsPi})
% \begin{align}

\begin{equation}
\begin{split}
   \alpha_1=&A d\varphi, \ \ \ \alpha_2=\sin^2\theta d\varphi, \\%  \ \ \ {\rm at}\ \ \theta=0, 
   \alpha'_1=&A' d\varphi', \ \ \ \alpha'_2=\sin^2\theta' d\varphi'.%,  \ \ \ {\rm at}\ \ \theta=\pi, 
   \end{split}
\end{equation}
% &
% &A' d\varphi', \ \ \ \sin^2\theta' d\varphi',  \ \ \ {\rm at}\ \ \theta'=\pi.}
% \end{align}
Clearly, they are smooth (analytic) in their domains including the poles $\theta=0$, and $\theta'=\pi$ respectively.
The remaining elements used in the definitions of the metric tensors $g$ and $g'$ are smooth (analytic) functions in their domains, also at the respective half-axes. 

In conclusion, the metric tensors $g$ and $g'$ give rise to a  metric tensor $\bar{g}$ uniquely defined on the manifold  constructed above, diffeomorphic to $\mathbb{R}\times S^3$, and admitting the U(1) bundle structure induced by the flow of the Killing vector field $\bar{\xi}$.   

% \color{violet}
For the construction above we have used one of the two possible choices (\ref{eq:b+}) or  (\ref{eq:b-}) of the parameter $b$.
Does the other vector have any special meaning in the resulting spacetime?
Does the outcome of this section depend on that choice?
To answer those questions suppose that 
\begin{equation*}
     b = b_+,
\end{equation*}
with the corresponding vector field $\xi$ renamed as $\xi_+$ for consistency.
Next, consider the other vector field $\xi_-$ corresponding to $b_-$.
Now we transform it to the coordinates adapted to $\xi_+$
\begin{equation*}
    \xi_- = \partial_\tau + (b_- -b_+)\partial_{\hat{\phi}} = \partial_\tau + \frac{P(\pi)}{2l}\partial_{\varphi}.
\end{equation*}
It is convenient to consider to consider a rescaled version of $\xi_-$
\begin{equation*}
     \hat{\xi}_-:=\partial_{{\varphi}} + \frac{2l}{P(\pi)}\partial_{\tau} .
\end{equation*}
Upon the triviality transformation (\ref{eq:t'}) it is expressed as 
\begin{equation*}
    \hat{\xi}_-=\partial_{{\varphi'}} - \frac{2l}{P(\pi)}\partial_{\tau'}. 
\end{equation*}

That symmetry indicates a special character of this vector field.
Indeed, we can introduce on every surface $r={\rm const}$ an auxiliary structure of the group SU(2), such that the vector field $\xi_+$ generates a left invariant vector field, the vector field $\xi_-$ a right invariant vector field, and the vector fields coincide at the circles corresponding to $\theta=0$, while they equal minus each other along the circles $\theta=\pi$.
{\color{black} For this reason the two points of intersection of orbits of $\xi_+$ and $\xi_-$ in Figures \ref{fig:globalPositiveLambda} and \ref{fig:globalLambdaNegative} are in fact two circles.} 
An important consequence of that symmetry between $\xi_+$ and $\xi_-$ is that the glued spacetime is independent of whether we chose $\xi_+$ or $\xi_-$ in order to define the generalised Misner gluing. 
{\color{black} Alternatively it is straightforward to explicitly check that after performing the gluing with $\xi_+$ it is possible to find four coordinate systems such that two of them are compatible (in the sense of (\ref{eq:killingCoords})) with $\xi_+$, one covering $\theta=0$ pole and the other $\theta=\pi$ pole, and the other two coordinate systems are an analogue for $\xi_-$.
Then the transformation between the coordinates compatible with $\xi_-$ satisfy precisely (\ref{eq:t'}) with $\sigma=-1$, thus showing that spacetimes constructed with either $\xi_+$ or $\xi_-$ arre equivalent.}

% {\color{blue} Finally we observe that although the decomposition ()\ref{eq:decomp}} is valid only when the norm of $\xi$ is non-zero, the metric tensors (\ref{eq:metricThroughHorizonKillingCoords0}) and (\ref{eq:metricThroughHorizonKillingCoordsPi}) do not distinguish the surfaces of $\xi^\mu\xi_\mu=0$ and are in fact well-posed thereon. This allows us to get rid of the above constraint and consider the whole spacetime as non-singular. 

\section{The global structure}
\label{sec:globalstructure}
The spacetime manifold is the entire $\mathbb{R}\times S^3$ provided $|l| > |a|$.  
Otherwise, if
\begin{equation}
\label{eq:l<a} 
|l| \le  |a|, 
\end{equation}
the vanishing of $\Sigma$ produces an non-removable singularity  at
\begin{equation*}
    (r,\theta)=(0,\theta_c)=(r',\theta'),
\end{equation*}
where the critical value $\theta_c$ is defined by
\begin{equation*}
    \cos\theta_c = -\frac{l}{a}. 
\end{equation*}
The type of the singularity can be characterised as a "ring" one, except for the case 
$$l =\pm a,$$
where the ring is shrunk to a point. 
% However, as in the Kerr spacetime, the surface 
% \begin{equation*}
%     r=0,\ \ \ \ \theta, \theta'\not=\theta_c,
% \end{equation*}
% which a curve has to cross in order to get from the $r>0$ to $r<0$ region, has a non-vanishing $3$-dimensional volume, hence it connects well the two spacetimes regions making a connected spacetime.
Notice however that, as in the Kerr spacetime, a curve going between the $r>0$ to $r<0$ regions has to cross the surface 
\begin{equation*}
    r=0,\ \ \ \ \theta, \theta'\not=\theta_c,
\end{equation*}
which has a non-vanishing $3$-dimensional volume, hence it connects the two spacetimes regions making a connected spacetime.

The vector field corresponding to $\partial_r$ in unprimed chart, and to $\partial_{r'}$ in the primed chart is globally defined and everywhere null
\begin{equation*}
    \bar{g}(\partial_r,\partial_r)=0.
\end{equation*}
A time orientation of spacetime can be defined by either declaring  
\begin{itemize}
    \item $\partial_r$ and $\partial_{r'}$ to be future directed, or
    \item $-\partial_r$ and $-\partial_{r'}$ to be future directed.
\end{itemize}
The coordinate transformation 
$$(\tau",r") := (-\tau, -r)   $$
maps the first case into the second (and vice versa), hence, without lack of generality we can assume that $ -\partial_r $ is a future pointing vector field. {\color{black}It should be emphasised that in doing so we allow for an arbitrary sign of the parameters $(a,l,m,\Lambda)$}. 

The Killing orbits are (generically) two dimensional surfaces endowed with the induced geometry
\begin{equation}
ds^2=-\frac{\mathcal{Q}}{\Sigma}((1-b A)d\tau-\frac{A}{P(0)} d\varphi)^2+\frac{P}{\Sigma}\sin^2\theta((a-b \rho ) d\tau-\frac{\rho}{P(0)} d\varphi)^2,
\end{equation}
The signature of the above is
\begin{itemize}
    \item $(-,+)$, if $\mathcal{Q}>0$,
    \item $(+,+)$, if $\mathcal{Q}<0$,
    \item $(0,+)$, i.e. null, if $\mathcal{Q}=0$.
\end{itemize}
If an orbit is timelike (or null) at a given point, than a vector field
\begin{equation}
\label{eq:futurePointinConstRVector}
  \eta=\partial_\tau + \frac{P(0)}{\rho(r_c)}(a-b \rho(r_c))\partial_\varphi,   
\end{equation}
where $r=r_c$ is constant, is timelike (or null).
Its time orientation is encoded by the scalar product
% $$ g(\partial_\tau + \frac{P(0)}{\rho(r_i)}(a-\rho(r_i)f)\partial_\varphi, -\partial_r) = -\frac{\Sigma}{\rho} < 0$$

\begin{equation*}
    g(\eta_c, -\partial_r) = -\frac{\Sigma}{\rho} < 0
\end{equation*}
hence it is always future pointing. 

Let $r_i$, $i=1,2,3,4$ be the roots of the polynomial $\mathcal{Q}$.
Then the surface of $r=r_i$ determines a Killing horizon (see \cite{LO2}) developed by the Killing vector $\eta_i$ with $r_c$ replaced with corresponding $r_i$.
% {\color{red}
% $$ r= r_i, \ \ \ i\in \{1,2,3,4\}.$$
% }
Similarly to the Kerr-(anti-) de Sitter space time, all of the roots cannot have the same sign. 
This follows from Viete's formulae: because there is no $r^3$ term in $\mathcal{Q}$ the roots must sum to $0$.
Another constraint is that $r_i=0$ corresponds to singularity and not a Killing horizon \cite{LO2}.
% {\color{red}It  is developed by the Killing vector 
% $$ \partial_\tau + \frac{P(0)}{\rho(r_i)}(a-\rho(r_i)f)\partial_\varphi. $$}
It also follows, that this Killing vector is always future pointing. 

In the very special case when
\begin{equation}
 \Lambda = \frac{3}{a^2+2l^2+2r^2_i},
\end{equation}
the coefficient at $\partial_\varphi$ of (\ref{eq:futurePointinConstRVector}) vanishes.
Then the horizon is developed by our Killing vector $\bar{\xi}$ itself. 
Hence, the null generators are closed and the space of the null generators is diffeomorphic to $S^2$. 
The geometry induced thereon is the limit of the metric tensor $^{(2)}q$ (\ref{eq:q2}), it is non-singular and smooth. That case was discovered and described in detail in \cite{LO2} along with their relation to solutions of Type D equation on Hopf bundle and isolated horizons \cite{localnohairPhysRevD.98.024008,hopf}

Another non-generic case is when the ratio of the coefficient at $\partial_\varphi$ to $4l/P(\pi)$ is rational.
Then the null generators will be finite and each such a case requires individual characterisation.
For a non-rational value of the ratio, all the null generators are infinite and each of them is dense in a 2-manifold contained in $S^3$.
The quotient space of those null generators is non-Hausdorff and lacks a differential structure.  

We can introduce a coordinate 
$$ \Omega := \frac{1}{r}$$
valid for either $r<0$ or $r>0$. The the metric tensor $g$ can be written as
\begin{equation*}
\begin{split}
g = \frac{1}{\Omega^2} \Bigg( & \frac{\Lambda}{3}\left((1-bA)d\tau -\frac{A}{P(0)}d\varphi\right)^2 + \frac{d\theta^2}{P}+ P \sin^2\theta\left(bd\tau +\frac{d\varphi}{P(0)}\right)^2+ \\
&-2d\Omega \left((1-bA)d\tau -\frac{A}{P(0)}d\varphi\right) +O(\Omega) \Bigg).
\end{split}
\end{equation*}
The surfaces of $ \Omega = 0$, corresponding to $r= \pm \infty$, define the future / past infinity equipped with an induced geometry
\begin{equation*}
   \frac{\Lambda}{3}\left((1-bA)d\tau -\frac{A}{P(0)}d\varphi\right)^2 + \frac{d\theta^2}{P}+ P \sin^2\theta\left(bd\tau +\frac{d\varphi}{P(0)}\right)^2  
\end{equation*}
of the signature depending on $\Lambda$ in the known way. 
Applying similar procedure ti the metric tensor (\ref{eq:metricThroughHorizonKillingCoordsPi}) one arrives at

\begin{equation*}
\begin{split}
g' = \frac{1}{\Omega^2} \Bigg(  &\frac{\Lambda}{3}\left((1-bA)d\tau' -\frac{A'}{P(\pi)}d\varphi'\right)^2 + \frac{d\theta'^2}{P}+ P \sin^2\theta'\left(bd\tau' +\frac{d\varphi'}{P(0)}\right)^2 \\
&-2d\Omega \left((1-bA)d\tau' -\frac{A'}{P(\pi)}d\varphi'\right) +O(\Omega) \Bigg),
\end{split}
\end{equation*}
which on the surfaces $\Omega=0$ gives the following geometry

\begin{equation*}
   \frac{\Lambda}{3}\left((1-bA)d\tau' -\frac{A'}{P(\pi)}d\varphi'\right)^2 + \frac{d\theta'^2}{P}+ P \sin^2\theta'\left(bd\tau' +\frac{d\varphi'}{P(\pi)}\right)^2. 
\end{equation*}
% \begin{equation}
%     \mathbb{R}\times S^3
% \end{equation}

\begin{figure}
    \centering
    \tdplotsetmaincoords{81}{0}
\begin{tikzpicture}[tdplot_main_coords, scale = 2]
\tdplotsetrotatedcoords{0}{0}{0};

\tikzmath{\rad=1;\h=3.5;\hcone=.08*\h;}
\coordinate (T) at (0,0,\h);
\coordinate (TR) at (\rad,0,\h);
\coordinate (TL) at (-\rad,0,\h);
\coordinate (B) at (0,0,0);
\coordinate (BR) at (\rad,0,0);
\coordinate (BL) at (-\rad,0,0);

\coordinate (TT) at (-1.2*\rad,0,\h*1.0);
\coordinate (BB) at (-1.2*\rad,0,-\h*0.0);

\coordinate (H4) at (\rad,0,0.8*\h);
\coordinate (H3) at (\rad,0,0.65*\h);
\coordinate (H2) at (\rad,0,0.5*\h);
\coordinate (H1) at (\rad,0,0.3*\h);

\coordinate (S3) at (1.0*\rad,0,-\h*0.12);
\draw[black,loosely dashed] (S3) arc (0:360:\rad);
\draw (S3) node[anchor=west,] {$S^3$};

\tikzmath{\fir=160;}
\coordinate (R1) at ({\rad*cos(\fir)},{\rad*sin(\fir)},0.33*\h);
\coordinate (R2) at ({\rad*cos(\fir)},{\rad*sin(\fir)},0.45*\h);
\draw [red,arrows={-latex[scale=2,red]}] (R1) -- (R2);
\draw (R2) node[anchor=north west,] {$-\partial_r$};

\coordinate (tauR) at (\rad,0,0.1*\h);
\coordinate (tauRm) at (\rad,0,0.05*\h);

\draw[solid, tdplot_rotated_coords, black] (T) circle (\rad);

%horyzonty i skraje
\draw[black] (BR) arc (0:-180:\rad);
\draw[black, dashed] (BR) arc (0:180:\rad);

\draw[blue] (H1) arc (0:-180:\rad);
\draw[blue, dashed] (H1) arc (0:180:\rad);
\draw[blue] (H2) arc (0:-180:\rad);
\draw[blue, dashed] (H2) arc (0:180:\rad);
\draw[blue] (H3) arc (0:-180:\rad);
\draw[blue, dashed] (H3) arc (0:180:\rad);
\draw[blue] (H4) arc (0:-180:\rad);
\draw[blue, dashed] (H4) arc (0:180:\rad);
\draw[ForestGreen] (tauR) arc (0:-180:\rad);
\draw[ForestGreen,dashed,arrows={-latex[scale=2,ForestGreen]}] (tauR) arc (0:-271:\rad);
\draw[ForestGreen,arrows={-latex[scale=2,ForestGreen]}] (tauR) arc (0:-80:\rad);
\draw[ForestGreen, dashed] (tauR) arc (0:180:\rad);

%podpisy
\draw (H1) node[anchor=west] {$H_1$};
\draw (H2) node[anchor=west] {$H_2$};
\draw (H3) node[anchor=west] {$H_3$};
\draw (H4) node[anchor=west] {$H_4$};
% \draw (TR) node[anchor=west] {$S^3$};
% \draw (BR) node[anchor=west] {$S^3$};
\draw (tauR) node[anchor=north west] {$\xi_-$};
\draw (tauR) node[anchor=south west] {$\xi_+$};
\draw (T) node[anchor=center] {$\mathcal{I}^+$};
\draw (B) node[anchor=center] {$\mathcal{I}^-$};
\draw (BB) node[anchor=south east] {$\mathbb{R}$};

%boki walca
\draw[black] (TL) -- (BL);
\draw[black] (TR) -- (BR);

\draw[loosely dashed, black, arrows={latex[scale=2,black]-}] (BB) -- (TT); %oś r

% stożki

\tikzmath{\fitau=80;\ficonepos=80;}
\coordinate (conetauA) at ({\rad*cos(\ficonepos)},{\rad*sin(\ficonepos)},{0.1*\h});
\tdplotsetrotatedcoordsorigin{(conetauA)};
\tdplotsetrotatedcoords{90-\ficonepos}{0}{0};
\draw[gray,fill=gray!60!,tdplot_rotated_coords,rotate around y={\fitau/2}] (0,0,{\hcone*cos(\fitau/2)}) circle ({\hcone*sin(\fitau/2)});
\draw[gray,fill=gray!60!,tdplot_rotated_coords] (0,0,0)--({\hcone*sin(\fitau)},0,{\hcone*cos(\fitau)})--(0,0,\hcone)--cycle;

\tikzmath{\fitau=70;\ficonepos=-100;}
\coordinate (conetauA) at ({\rad*cos(\ficonepos)},{\rad*sin(\ficonepos)},{0.1*\h});
\tdplotsetrotatedcoordsorigin{(conetauA)};
\tdplotsetrotatedcoords{90-\ficonepos}{0}{0};
\draw[gray,fill=gray!60!,tdplot_rotated_coords] (0,0,0)--({\hcone*sin(\fitau)},0,{\hcone*cos(\fitau)})--(0,0,\hcone)--cycle;
\draw[gray,fill=gray!60!,tdplot_rotated_coords,rotate around y={\fitau/2}] (0,0,{\hcone*cos(\fitau/2)}) circle ({\hcone*sin(\fitau/2)});

\tikzmath{\fitau=70;\ficonepos=-105;}
\coordinate (conetauA) at ({\rad*cos(\ficonepos)},{\rad*sin(\ficonepos)},{0.0*\h});
\tdplotsetrotatedcoordsorigin{(conetauA)};
\tdplotsetrotatedcoords{90-\ficonepos}{0}{0};
\draw[gray,fill=gray!60!,tdplot_rotated_coords] (0,0,0)--({\hcone*sin(\fitau)},0,{\hcone*cos(\fitau)})--(0,0,\hcone)--cycle;
\draw[gray,fill=gray!60!,tdplot_rotated_coords,rotate around y={\fitau/2}] (0,0,{\hcone*cos(\fitau/2)}) circle ({\hcone*sin(\fitau/2)});

\tikzmath{\fitau=90;\ficonepos=85;}
\coordinate (conetauA) at ({\rad*cos(\ficonepos)},{\rad*sin(\ficonepos)},{0.3*\h});
\tdplotsetrotatedcoordsorigin{(conetauA)};
\tdplotsetrotatedcoords{90-\ficonepos}{0}{0};
\draw[gray,fill=gray!60!,tdplot_rotated_coords,rotate around y={\fitau/2}] (0,0,{\hcone*cos(\fitau/2)}) circle ({\hcone*sin(\fitau/2)});
\draw[gray,fill=gray!60!,tdplot_rotated_coords] (0,0,0)--({\hcone*sin(\fitau)},0,{\hcone*cos(\fitau)})--(0,0,\hcone)--cycle;

\tikzmath{\fitau=90;\ficonepos=-90;}
\coordinate (conetauA) at ({\rad*cos(\ficonepos)},{\rad*sin(\ficonepos)},{0.3*\h});
\tdplotsetrotatedcoordsorigin{(conetauA)};
\tdplotsetrotatedcoords{90-\ficonepos}{0}{0};
\draw[gray,fill=gray!60!,tdplot_rotated_coords] (0,0,0)--({\hcone*sin(\fitau)},0,{\hcone*cos(\fitau)})--(0,0,\hcone)--cycle;
\draw[gray,fill=gray!60!,tdplot_rotated_coords,rotate around y={\fitau/2}] (0,0,{\hcone*cos(\fitau/2)}) circle ({\hcone*sin(\fitau/2)});

\tikzmath{\d=0.1*\h;\rminus=sqrt(\rad*\rad+\d*\d*0.25);\al=atan(\d/2/\rad);}
% \coordinate (tauCenter) at (0,0,0.1*\h+\d/2);
% \tdplotsetrotatedcoordsorigin{(tauCenter)};
% \tdplotsetrotatedcoords{0}{\al}{0};
% \draw[orange, tdplot_rotated_coords] (0,0,0) circle (\rminus);

\tdplotsetrotatedcoordsorigin{(tauRm)};
\tdplotsetrotatedcoords{0}{\al}{0};

\draw[orange,tdplot_rotated_coords] (0,0,0) arc (0:-180:\rad);
\draw[orange,tdplot_rotated_coords,dashed,arrows={-latex[scale=2,orange]}] (0,0,0) arc (0:-271:\rad);
\draw[orange,tdplot_rotated_coords,arrows={-latex[scale=2,orange]}] (0,0,0) arc (0:-80:\rad);
\draw[orange,tdplot_rotated_coords, dashed] (0,0,0) arc (0:180:\rad);

\end{tikzpicture}
    \caption{Global structure of the spacetime for $\Lambda>0$. Each point on the cylinder corresponds to a $S^2$. Each circle corresponds to a $S^3$ although via different sections. Four uppermost, blue circles $H_i$ correspond to surfaces of constant $r=r_i$ for $i=1,2,3,4$, i.e. Killing horizons. Green, lowermost circle is $S^3$ projected along the orbit of $\xi_+$. The orbit is closed and spacelike near both of the $\mathcal{I}$. Similarly the orbit of $\xi_-$ is shown orange. Points of intersection of the orbits are circles. Future oriented (in the direction of $\partial_{-r}$) parts of lightcones are shown. }
    \label{fig:globalPositiveLambda}
\end{figure}

\begin{figure}
    \centering
    \tdplotsetmaincoords{90}{9}
\begin{tikzpicture}[tdplot_main_coords, scale = 2,rotate around y =90]
% \fill[color = orange, opacity = 1] (0,0,0) circle (1cm);
% \fill[color = orange, opacity = 1] (0.52cm,0.85cm) -- (-0.52cm,0.85cm) -- (0cm,1.1cm) -- cycle;
% \fill[color = orange, opacity = 1] (0.52cm,-0.85cm) -- (-0.52cm,-0.85cm) -- (0cm,-1.1cm) -- cycle;
\tdplotsetrotatedcoords{0}{0}{0};

\tikzmath{\rad=1;\h=3.5;\hcone=.08*\h;}
\coordinate (T) at (0,0,\h);
\coordinate (TR) at (\rad,0,\h);
\coordinate (TL) at (-\rad,0,\h);
\coordinate (B) at (0,0,0);
\coordinate (BR) at (\rad,0,0);
\coordinate (BL) at (-\rad,0,0);

\coordinate (TT) at (1.3*\rad,0,\h*1.0);
\coordinate (BB) at (1.3*\rad,0,-\h*0.0);

\coordinate (H4) at (\rad,0,0.8*\h);
\coordinate (H3) at (\rad,0,0.65*\h);
\coordinate (H2) at (\rad,0,0.5*\h);
\coordinate (H1) at (\rad,0,0.3*\h);

\coordinate (tauR) at (\rad,0,0.1*\h);
\coordinate (tauRm) at (\rad,0,0.05*\h);

\draw[solid, black] (T) circle (\rad);

%horyzonty i skraje
\draw[black] (BR) arc (0:-180:\rad);
\draw[black, dashed] (BR) arc (0:180:\rad);

\draw[blue] (H1) arc (0:-180:\rad);
\draw[blue, dashed] (H1) arc (0:180:\rad);
\draw[blue] (H2) arc (0:-180:\rad);
\draw[blue, dashed] (H2) arc (0:180:\rad);
\draw[blue] (H3) arc (0:-180:\rad);
\draw[blue, dashed] (H3) arc (0:180:\rad);
\draw[blue] (H4) arc (0:-180:\rad);
\draw[blue, dashed] (H4) arc (0:180:\rad);
\draw[ForestGreen] (tauR) arc (0:-180:\rad);
\draw[ForestGreen,dashed,arrows={-latex[scale=2,ForestGreen]}] (tauR) arc (0:-271:\rad);
\draw[ForestGreen,arrows={-latex[scale=2,ForestGreen]}] (tauR) arc (0:-80:\rad);
\draw[ForestGreen, dashed] (tauR) arc (0:180:\rad);

%podpisy
\draw (H1) node[anchor=north] {$H_1$};
\draw (H2) node[anchor=north] {$H_2$};
\draw (H3) node[anchor=north] {$H_3$};
\draw (H4) node[anchor=north] {$H_4$};
% \draw (TR) node[anchor=north] {$S^3$};
% \draw (BR) node[anchor=north] {$S^3$};
\draw (tauR) node[anchor=north east] {$\xi_-$};
\draw (tauR) node[anchor=north west] {$\xi_+$};
\draw (T) node[anchor=center] {$\mathcal{I}^+$};
\draw (B) node[anchor=center] {$\mathcal{I}^-$};
% \draw (TT) node[anchor=north west] {$-\partial_{r}$};
\draw (BB) node[anchor=north east] {$\mathbb{R}$};

%boki walca
\draw[black] (TL) -- (BL);
\draw[black] (TR) -- (BR);

\draw[loosely dashed, black, arrows={latex[scale=2,black]-}] (BB) -- (TT); %oś r

\tikzmath{\fir=20;}
\coordinate (R1) at ({\rad*cos(\fir)},{\rad*sin(\fir)},0.33*\h);
\coordinate (R2) at ({\rad*cos(\fir)},{\rad*sin(\fir)},0.45*\h);
\draw [red,arrows={-latex[scale=2,red]}] (R1) -- (R2);
\draw (R2) node[anchor=south east,] {$-\partial_r$};

\coordinate (S3) at (1.0*\rad,0,-\h*0.15);
\draw[black,loosely dashed] (S3) arc (0:360:\rad);
\draw (S3) node[anchor=east] {$S^3$};

% stożki

\tikzmath{\fitau=120;\ficonepos=80;}
\coordinate (conetauA) at ({\rad*cos(\ficonepos)},{\rad*sin(\ficonepos)},{0.1*\h});

\tdplotsetrotatedcoordsorigin{(conetauA)};
\tdplotsetrotatedcoords{90-\ficonepos}{90}{0};
% \draw[red,fill,tdplot_rotated_coords] (conetauA) circle (0.1);
\draw[gray,fill=gray!60!,tdplot_rotated_coords,rotate around y={\fitau/2}] (0,0,{\hcone*cos(\fitau/2)}) circle ({\hcone*sin(\fitau/2)});
\draw[gray,fill=gray!60!,tdplot_rotated_coords] (0,0,0)--({\hcone*sin(\fitau)},0,{\hcone*cos(\fitau)})--(0,0,\hcone)--cycle;

\tikzmath{\fitau=120;\ficonepos=-100;}
\coordinate (conetauA) at ({\rad*cos(\ficonepos)},{\rad*sin(\ficonepos)},{0.1*\h});
\tdplotsetrotatedcoordsorigin{(conetauA)};
\tdplotsetrotatedcoords{-(90-\ficonepos)}{-90}{0};
\draw[gray,fill=gray!60!,tdplot_rotated_coords] (0,0,0)--({\hcone*sin(\fitau)},0,{\hcone*cos(\fitau)})--(0,0,\hcone)--cycle;
\draw[gray,fill=gray!60!,tdplot_rotated_coords,rotate around y={\fitau/2}] (0,0,{\hcone*cos(\fitau/2)}) circle ({\hcone*sin(\fitau/2)});

\tikzmath{\fitau=120;\ficonepos=-95;}
\coordinate (conetauA) at ({\rad*cos(\ficonepos)},{\rad*sin(\ficonepos)},{0.0*\h});
\tdplotsetrotatedcoordsorigin{(conetauA)};
\tdplotsetrotatedcoords{-(90-\ficonepos)}{-90}{0};
\draw[gray,fill=gray!60!,tdplot_rotated_coords] (0,0,0)--({\hcone*sin(\fitau)},0,{\hcone*cos(\fitau)})--(0,0,\hcone)--cycle;
\draw[gray,fill=gray!60!,tdplot_rotated_coords,rotate around y={\fitau/2}] (0,0,{\hcone*cos(\fitau/2)}) circle ({\hcone*sin(\fitau/2)});

\tikzmath{\fitau=90;\ficonepos=85;}
\coordinate (conetauA) at ({\rad*cos(\ficonepos)},{\rad*sin(\ficonepos)},{0.3*\h});
\tdplotsetrotatedcoordsorigin{(conetauA)};
\tdplotsetrotatedcoords{90-\ficonepos}{90}{0};
\draw[gray,fill=gray!60!,tdplot_rotated_coords,rotate around y={\fitau/2}] (0,0,{\hcone*cos(\fitau/2)}) circle ({\hcone*sin(\fitau/2)});
\draw[gray,fill=gray!60!,tdplot_rotated_coords] (0,0,0)--({\hcone*sin(\fitau)},0,{\hcone*cos(\fitau)})--(0,0,\hcone)--cycle;

\tikzmath{\fitau=90;\ficonepos=-90;}
\coordinate (conetauA) at ({\rad*cos(\ficonepos)},{\rad*sin(\ficonepos)},{0.3*\h});
\tdplotsetrotatedcoordsorigin{(conetauA)};
\tdplotsetrotatedcoords{-(90-\ficonepos)}{-90}{0};
\draw[gray,fill=gray!60!,tdplot_rotated_coords] (0,0,0)--({\hcone*sin(\fitau)},0,{\hcone*cos(\fitau)})--(0,0,\hcone)--cycle;
\draw[gray,fill=gray!60!,tdplot_rotated_coords,rotate around y={\fitau/2}] (0,0,{\hcone*cos(\fitau/2)}) circle ({\hcone*sin(\fitau/2)});

\tikzmath{\d=0.1*\h;\rminus=sqrt(\rad*\rad+\d*\d*0.25);\al=atan(\d/2/\rad);}
% \coordinate (tauCenter) at (0,0,0.1*\h+\d/2);
% \tdplotsetrotatedcoordsorigin{(tauCenter)};
% \tdplotsetrotatedcoords{0}{90+\al}{0};
% \draw[orange, tdplot_rotated_coords] (0,0,0) circle (\rminus);
\tdplotsetrotatedcoordsorigin{(tauRm)};
\tdplotsetrotatedcoords{0}{90+\al}{0};
\draw[orange,tdplot_rotated_coords] (0,0,0) arc (0:-180:\rad);
\draw[orange,tdplot_rotated_coords,dashed,arrows={-latex[scale=2,orange]}] (0,0,0) arc (0:-271:\rad);
\draw[orange,tdplot_rotated_coords,arrows={-latex[scale=2,orange]}] (0,0,0) arc (0:-80:\rad);
\draw[orange,tdplot_rotated_coords, dashed] (0,0,0) arc (0:180:\rad);

\end{tikzpicture}
        \caption{Global structure of the spacetime for $\Lambda<0$. Each point on the cylinder corresponds to a $S^2$. Each circle corresponds to a $S^3$ although via different sections. Four rightmost, blue circles $H_i$ correspond to surfaces of constant $r=r_i$ for $i=1,2,3,4$, i.e. Killing horizons. Green, leftmost circle is $S^3$ projected along the orbit of $\xi$. The orbit is manifestly closed and timelike near both of the $\mathcal{I}$. Similarly the orbit of $\xi_-$ is shown orange. Points of intersection of the orbits are circles. Future oriented (in the direction of $\partial_{-r}$) parts of lightcones are shown. }
    \label{fig:globalLambdaNegative}
\end{figure}

There are still two discrete  degrees of freedom we have not discussed yet.
The first one is the causal orientation.
The second would be using the outgoing Eddington-Finkelstein coordinates 
\begin{equation}
    dv:=dt-\frac{\rho}{\mathcal{Q}}dr,\quad
    d\Tilde{\phi}:=d\phi +\frac{a}{\mathcal{Q}}dr.
\end{equation}
rather than incoming ones (\ref{eq:EddigntonCoords}).
However, the latter is equivalent to $(v,a,l)\xrightarrow{}(-v,-a,-l)$. 
Of course, there is still  the symmetry of reversing signs of both: the $r$ coordinate and mass $(r,m)\xrightarrow[]{}(-r-m)$.

The schematics of global structure summarizing the above constructions and conventions is shown in Figures \ref{fig:globalPositiveLambda} and \ref{fig:globalLambdaNegative} for positive and negative values of cosmological constant, respectively.

\section{Summary} 
The result of this paper is a $4$ dimensional  family (parametrised by the quadruple $(m,a,l,\Lambda)$) of globally defined spacetimes that are locally isometric to the Kerr-NUT-(anti) de Sitter metric tensors (\ref{eq:KNdS-metric}), however, they do not suffer the singularities along the axis $\theta=0$ and $\theta=\pi$.
The spacetime manifold is obtained by gluing the manifolds  
\begin{equation}
    S_1\times \mathbb{R}\times \left(S^2\setminus \{p_\pi\}\right)
\end{equation}
parametrised by $(\tau,r,\theta,\varphi)$,
and
\begin{equation}
\label{bundle4"} S_1\times \mathbb{R}\times \left(S^2\setminus \{p_0\}\right)
\end{equation}
parametrised by $(\tau',r',\theta',\varphi')$, together with the transformation  (\ref{eq:t'}).
The coordinates $(\theta,\varphi)$ and $(\theta',\varphi')$ are the standard spherical coordinates on $S^2$, while the variables $\tau$ and $\tau'$ parametrise  circles.
The points $p_0$ and $p_\pi$ are the poles of $S^2$ corresponding $\theta=0$ and $\theta'=\pi$.  

For every choice of the parameters $(m,a,l,\Lambda)$, the spacetime metric tensor is defined  by (\ref{eq:metricThroughHorizonKillingCoords0}) and (\ref{eq:metricThroughHorizonKillingCoordsPi}), with the functions $\mathcal{Q}$, $\Sigma$, $\rho$, $P>0$ and $A$ defined by (\ref{eq:KNdS-metric}).
The only possible singularity of our spacetime metric tensor may be caused (depending on the ratio $l/a$, see (\ref{eq:l<a})) by vanishing of the function $\Sigma$.
The singularity has a similar character to that of Kerr spacetime - in particular it does not restrict the domain of the $r$ coordinate $-\infty <r<\infty$. 
What is new about our result, is the simultaneous presence of  the Kerr parameter $a\not=0$, the NUT parameter $l\not=0$ and the cosmological constant $\Lambda\not = 0$. 

An underlying structure for the construction was the assumed isometric action of the group U(1) that makes the spacetime  a principal fiber bundle 
$$ \mathbb{R}\times S^3\ \rightarrow\ \mathbb{R}\times S^2$$
(modulo the possible singularities discussed above).  
The key element of our method  was determining a suitable candidate Killing vector field in the the Kerr-NUT-(anti) de Sitter metric tensor (\ref{eq:KNdS-metric}) that could become a generator of that non-singular action of U(1) on $\mathbb{R}\times S^3$.
We have achieved that by studying  the geometry of the spaces of non-null orbits of each Killing vector field of the Kerr-NUT-(Anti) de Sitter spacetime, and selecting those that induce non-singular $3$-geometry.     

We studied the global structure of the constructed spacetimes.
Depending on the value of the cosmological constant $\Lambda$, our spacetime is asymptotically  de Sitter or anti-de Sitter.  
We derive the conformal geometry of the conformal infinity and find it is non-singular as well {\color{black} (topologically, the conformal infinities are two copies of $3$-sphere)}.  In particular, in the case $\Lambda>0$ the signature of the infinity is $(+++)$ and a spacetime neighborhood seems to be hyperbolic.
The spacetime contains up to four Killing horizons corresponding to the roots of the function $\mathcal{Q}$.
Generically, the null generators of the horizons are infinite curves and each of them densely covers a $2$-surface. % diffeomorphic to torus.
Hence the space of the null generators is not a differentiable $2$-dimensional manifold.
For special values of $(m,a,l,\Lambda)$ (see (\ref{eq:projectivehor})) the null generators of one of the horizons coincide with the fibers of the bundle (\ref{bundle4"}).
Then, the horizon is projectively non-singular, that is the space of the null generators has a non-singular geometry diffeomorphic to $S^2$ \cite{LO2}.

\textit{Acknowledgements} 
We wish to thank Piotr Chruściel, Lionel Mason, Reinhard Meinel and Sir Roger Penrose for useful comments and fruitful discussions of our work.
Special gratitude is due for Wojciech Kamiński for his help in proving the  smoothness of our metrics.
This work was partially supported by the Polish National Science Centre
grants No. 2017/27/B/ST2/02806 and No. 2016/23/P/ST1/04195.

\color{black}
\bibliographystyle{apsrev4-2}
\bibliography{bibliography}

\end{document}